\documentclass[preprint,showpacs,preprintnumbers,amsmath,amssymb,superscriptaddress, nofootinbib]{revtex4}  
\usepackage{graphicx,color}
\usepackage{amsmath,amssymb}
\usepackage{url}
\usepackage{epstopdf}
\usepackage{slashed}

\newcommand{\be}{\begin{equation}}
\newcommand{\ee}{\end{equation}}
\newcommand{\bea}{\begin{eqnarray}}
\newcommand{\eea}{\end{eqnarray}}

\newcommand{\crn}{\nonumber \\}

\newcommand{\fr}{\frac}

\newcommand{\bc}{\begin{center}}
	\newcommand{\ec}{\end{center}}

\newcommand {\ba}{\begin{array}}
	\newcommand {\ea}{\end{array}}
\newcommand{\ben}{\begin{enumerate}}
	\newcommand{\een}{\end{enumerate}}
	
\usepackage{epsfig,graphicx} 
\usepackage{bm}
\usepackage{dcolumn}
\begin{document}

\title{ LFV decays in a 3-3-1 model with singlet leptoquarks}

\author{N.H.T. Nha}\email{nguyenhuathanhnha@vlu.edu.vn}
\affiliation{Subatomic Physics Research Group, Science and Technology Advanced Institute, Van Lang University, Ho Chi Minh City, Vietnam}
\affiliation{Faculty of Applied Technology, Van Lang School of Technology, Van Lang University, Ho Chi Minh City, Vietnam}

\author{L.T. Hue}\email{lethohue@vlu.edu.vn}
\affiliation{Subatomic Physics Research Group, Science and Technology Advanced Institute, Van Lang University, Ho Chi Minh City, Vietnam}
\affiliation{Faculty of Applied Technology, Van Lang School of Technology, Van Lang University, Ho Chi Minh City, Vietnam}
\author{N. T. K. Ngan }\email{ntkngan@ctu.edu.vn}
\affiliation{Department of Physics, Can Tho University,
	3/2 Street, Can Tho, Vietnam}
\author{P.T. Bich}\email{ptbich@nctu.edu.vn}
\affiliation{Nam Can Tho University, 168 Nguyen Van Cu Street, An Binh Ward, Can Tho City, Vietnam}
\author{T.T. Hong}\email{tthong@agu.edu.vn}
\affiliation{{An Giang University,  An Giang, Vietnam}} 
\affiliation{{Vietnam National University, Ho Chi Minh City, Vietnam}}
\author{N.T. Tham \footnote{corresponding author}} 
\email{nguyenthitham@hpu2.edu.vn}
\affiliation{Department of Physics, Hanoi Pedagogical University 2,
 Phu Tho, Vietnam}

\begin{abstract}
Motivated by a recent study of the 3-3-1 model supplemented with a singlet scalar leptoquark, which successfully explains the muon anomalous magnetic moment and the decay $\mu\to e\gamma$ within current experimental constraints, we extend the phenomenological analysis of this framework to include the lepton-flavor-violating decays of the Standard Model-like Higgs and the $Z$ boson. An interesting feature is that the branching ratios of these decays exhibit a nearly linear correlation with the corresponding charged lepton flavor-violating radiative decays, namely $\mathrm{Br}(h,Z\to e_b^\pm e_a^\mp)\propto\mathrm{Br}(e_b\to e_a\gamma)$. Furthermore, the model exhibits a complementary interplay between the charged-lepton anomalous magnetic moments: parameter regions with $|\Delta a_{\mu}|\geq10^{-11}$ favor $\mathrm{Br}(h,Z\to\tau^\pm\mu^\mp)$ reaching their current experimental upper limits while keeping $|\Delta a_e|$ negligible, whereas regions with $10^{-14}\leq|\Delta a_e|\leq 8\times10^{-13}$ instead allow $\mathrm{Br}(h,Z\to\tau^\pm e^\mp)$ to approach the present experimental sensitivities but simultaneously suppress $|\Delta a_\mu|$ to the level of $\mathcal{O}(10^{-16})$.
\end{abstract}


\maketitle
\section{\label{intro} Introduction}
\allowdisplaybreaks
The 3-3-1 models were constructed based on the gauge group $SU(3)_C\otimes SU(3)_L \otimes U(1)_X$ \cite{Singer:1980sw, Pleitez:1992xh, Ozer:1995xi, Foot:1994ym, Frampton:1992wt}, providing an interesting explanation for the existence of three fermion families based on the anomaly-free requirements in the fermion sector \cite{Frampton:1992wt}. In particular, the cancellation of gauge anomalies requires that the number of fermion generations be a multiple of the number of color quanta, which naturally leads to three families. In addition, the non-universal assignment of quark representations among different generations offers a possible explanation for the large mass of the top quark as well as phenomenological signatures Beyond the Standard Model (BSM). Among various realizations, the 3-3-1 models with right-handed neutrinos (331RHN) have attracted significant attention due to their rich phenomenology in both the gauge and scalar sectors.  { Apart from predicting new gauge bosons and exotic fermions at the TeV scale, these models accommodate active neutrino masses and mixings  along with lepton-flavor-violating (LFV) processes as promoting signals of  new physics \cite{Cogollo:2008zc, Catano:2012kw, Dias:2012xp, Boucenna:2015zwa, deSousaPires:2018fnl,  Hue:2021xap, Hong:2022xgx, Escalona:2025jla}.}

On the other hand, the recent update discrepancy between the Standard Model (SM) prediction and the experimental measurements of the anomalous magnetic moments (AMMs) of charged leptons, especially the muon $(g-2)_{\mu}$ \cite{Muong-2:2023cdq, Muong-2:2025xyk}, still provides a strong motivation to explore new physics BSM. {In 2025, this discrepancy from experimental collaborations and theory of the muon continues to indicate that a deviation within $1\sigma$ can reach the value of $10^{-9}$ \cite{Aliberti:2025beg, Li:2025myw}. Namely, defining that $a_{e_a}\equiv (g-2)_{e_a}$, the deviation is \cite{Aliberti:2025beg}:}
\begin{align}
\label{eq:damu}
\Delta a^{\mathrm{new}}_{\mu} &\equiv  a^{\mathrm{exp}}_{\mu} -a^{\mathrm{SM}}_{\mu} =\left(3.8\pm 6.3 \right) \times 10^{-10} .
\end{align}
{In addition, possible deviations between experimental measurements \cite{Hanneke:2008tm, Parker:2018vye, Morel:2020dww, Fan:2022eto} and  the SM prediction for  the electron AMM   further suggest that a unified framework addressing lepton-flavor is highly desirable.} A generic feature of many BSM scenarios is that the same new interactions contributing to AMMs may also induce charged LFV (cLFV) decays, $e_b \to e_a \gamma$, as well as LFV decays of the SM-like Higgs boson (LFV$h$) and neutral gauge boson Z (LFV$Z$). Consequently, the stringent experimental bounds on cLFV decays, for instance, future sensitivity in 2026 of $\text{Br}(\mu \to e\gamma) < 6\times 10^{-14}$ \cite{MEGII:2018kmf, Belle-II:2018jsg, MEGII:2025gzr}, impose severe constraints on the parameter space of models attempting to explain the $(g-2)_{a_e}$ AMMs. This strong interplay between AMM observables and these upper bounds of LFV processes has represented a key challenge for model building.

In recent years, extensions of the 3-3-1 framework with additional scalar or fermionic degrees of freedom \cite{Hong:2022xjg,Hong:2024yhk,Hong:2024swk,Hieu:2025kxt} have been proposed to simultaneously address {the sizable one-loop contributions to $(g-2)_{e_a}$ anomalies and LFV observables, along with the improvements of relevant experimental data. In particular, models involving leptoquarks have gained considerable attention \cite{Cheung:2001ip, Doff:2024cap, Mahanta:2001yc}, since they naturally couple leptons and quarks and can induce sizable chirality-flipping effects at the loop level, which are essential for enhancing $\Delta a_{\mu}$ up to the level of $\mathcal{O}(10^{-9})$ \cite{Doff:2024cap}.} However, most of the existing analyses focus on specific channels instead of including all relevant LFV processes. In particular, a systematic evaluation of all one-loop contributions, including both fermionic and scalar, is necessary in order to fully capture the correlations among different observables and to reliably determine the allowed parameter space.

Therefore, in this work, we extend previous studies by performing a general and complete analysis of LFV processes and AMM of charged leptons in the 3-3-1 model supplemented by a singlet scalar leptoquark (called 331LQ for short) \cite{Doff:2024cap}.
This general framework not only reproduces the known results in specific limits but also reveals new correlations between observables that have not been fully explored in previous studies. Furthermore, we perform a detailed numerical analysis to identify the regions of parameter space {that are consistent with current experimental data on various LFV processes and $(g-2)_{e_a}$, focusing on interesting regions that allow sizable values of  $\Delta a_{e,\mu}$.} Special attention is paid to the correlations between the branching ratios (Brs) of LFV$h$ and cLFV decays. {In particular, we address two key questions: whether the 331LQ framework can simultaneously accommodate large values of $\Delta a_{e}$ and $\Delta a_{\mu}$, and which LFV decay channels are most likely to be observed experimentally, thereby providing the strongest constraints on the viable parameter space.}

The paper is organized as follows. In Sec.~\ref{model}, {we review the structure of the 331LQ model.} In Sec.~\ref{sec_coupling}, we present the relevant interactions and derive the general analytical expressions for the one-loop contributions to LFV processes and {AMMs.} In Sec.~\ref{sec_numerical}, we perform a comprehensive numerical analysis and discuss the resulting phenomenological implications. Finally, our conclusions are given in Sec.~\ref{conclusion}. Furthermore, in Appendix.~\ref{app:higgs}, we will summarize all of the calculations relevant to the Higgs potential, leptoquark mass, and the triple coupling of the SM-like Higgs boson with leptoquarks appearing in this model. 

\section{\label{model} The 3-3-1 model with singlet leptoquark}
\subsection{Particle content and neutrino masses from the ISS mechanism}
The 331LQ model is constructed by adding the new singlet leptoquark $S$ into the original 3-3-1 model, where $S$ can interact with both leptons and quarks. In the leptonic sector, the left-handed leptons are represented as a triplet by $SU(3)_L$, while right-handed leptons are a singlets of this gauge group \cite{Doff:2024cap, Doff:2006rt}, namely
\begin{equation}
L_{a}= \begin{pmatrix}
\nu_a \\
e_a \\
 (\nu_a)^c
\end{pmatrix}_L  \sim (1,3,-1/3), \; e_{aR} \sim (1,1,-1),
\end{equation}
with $a=1,2,3$ corresponding to three SM lepton generations. 

In the hadronic sector, we chose to represent quarks with different flavors corresponding to the triplet and anti-triplet by $SU(3)_L$ gauge group as follows
\begin{align}
Q_{kL} =& (d_k, -u_k, D_k)_L^T \sim (3, \overline{3}, 0), \;D_{kR} \sim (3,1,-1/3),\crn
Q_{3L} =& (u_3, d_3, U_3)_L^T \sim (3, 3, 1/3), \; U_{3R} \sim (3,1,2/3),\crn
d_{aR}\sim& (3,1,-1/3), \; u_{aR} \sim (3,1,2/3),
\end{align}
where $k=1,2$ is restricted to only first two generations.

The three Higgs triplets of the model are $\rho=(\rho^+_1, \rho^0, \rho^+_2)^T \sim (1,3,2/3)$, $\eta=(\eta_1^0, \eta^-, \eta^0_2)^T \sim (1,3,-1/3)$, $\chi=(\chi_1^0, \chi^-, \chi^0_2)^T\sim (1,3,-1/3)$,  and the {new charged leptoquark} $S \sim (3,1,1/3)$. Besides, the electric charge operator is defined by the gauge group $SU(3)_L \otimes U(1)_X$: $Q = T_3 - T_8/\sqrt{3} +X$ \cite{Foot:1994ym}. Consequently, we determined the electric charge of the new leptoquark $S$ with $Q(S)=1/3$. Furthermore, all quark and lepton masses at tree-level are generated by the vacuum expectation values (vev): $\langle\rho \rangle=(0,\,\frac{v_\rho}{\sqrt{2}},\,0)^T$, $\langle \eta \rangle=(\frac{v_\eta}{\sqrt{2}},\,0,\,0)^T$ and $\langle \chi \rangle=(0,\,0,\,\frac{v_\chi}{\sqrt{2}})^T$.

\subsection{Mass matrices and Yukawa interactions}

The Lagrangian Yukawa interactions generating tree level masses for all quarks and leptons in the model \cite{Chang:2006aa, Okada:2016whh, Doff:2024cap} as follows
\begin{align}\label{Lmass}
- \mathcal{L}_Y= &
g^d_{ka}\overline{Q}_{kL}\eta^{*}d_{aR} + g^d_{3a}\overline{Q}_{3L}\rho d_{aR}  + g^u_{ka}\overline{Q}_{kL}\rho^{*}u_{aR} + g^u_{3a}\overline{Q}_{3L}\eta u_{aR}  + y^{\nu}_{ab} \overline{(L_{aL})^c} L_{bL}\rho + y^e_{ab}\overline{L_{aL}}\rho e_{bR}
\crn& +g^D_{ka}\overline{Q}_{kL}\chi^{*}D_{kR} + g^U_{33}\overline{Q}_{3L}\chi U_{3R}  +\mathrm{h.c.},
\end{align}
where all sums are taken over $k = 1, 2$, and $a = 1,2,3$. We note that Lagrangian in Eq. \eqref{Lmass} respect a $Z_2$ discrete symmetry introduced Ref. \cite{Okada:2016whh} so that the SM quarks do not mix with exotic ones. This still allows to generate masses and mixing of quarks consistent with experiments, namely the Lagrangian and mass matrices of SM -like quark are:
\begin{align}
\label{eq:Lqmass}
-\mathcal{L}^q_{\mathrm{mass}}= &\sum_{f=u,d}\overline{q_{fL}}\mathcal{M}_fq_{fR} +\mathrm{h.c.},
\crn \mathcal{M}_u=& \frac{v_{\rho}}{\sqrt{2}}\begin{pmatrix}
-g^u_{11}& -g^u_{12} & -g^u_{13} \\ 
-g^u_{21}& -g^u_{22} & -g^u_{23} \\ 
g^u_{31}t_{\beta}& g^u_{32}t_{\beta} & g^u_{33} t_{\beta}  
\end{pmatrix} , \; 
\mathcal{M}_d= \frac{v_{\rho}}{\sqrt{2}}\begin{pmatrix}
g^d_{11}t_{\beta}& g^d_{12}t_{\beta} & g^d_{13} t_{\beta} \\ 
g^d_{21}t_{\beta}& g^d_{22}t_{\beta} & g^d_{23} t_{\beta} \\
g^d_{31}&g^d_{32} &g^d_{33} 
\end{pmatrix} ,
\end{align}
where $q_{fL(R)}=(f_1,f_2,f_3)^T_{L(R)}$, and 
\begin{align}
\label{eq:vtb}
v=\sqrt{v_{\eta}^2+v_{\rho}^2},\;  t_\beta = \frac{v_\eta}{v_\rho}.
\end{align}
We note that {the specific case of   $t_\beta = 1$ mentioned in  Ref.~\cite{Doff:2024cap, DeJesus:2020yqx}. It can be seen here that $t_{\beta}$ does not significantly affect the LFV decay amplitudes. }

In general, assuming that the flavor and mass states of left-handed quarks are not the same but relate to each others through two unitary transformations: $q_{fL}\equiv (f_1,f_2,f_3)^T=V^{f\dagger}_{L}\hat{q}_{fL}$  with $f=u,d$ relating to the two following mass base:   $\hat{q}_{dL}=(d,s,b)^T_L$ and $ \hat{q}_{uL}=(u,c,t)^T_L$. Correspondingly, the transformations between the diagonal mass matrices and the original ones are: $V^f_{L}\mathcal{M}_fV^{f\dagger}_R= \hat{\mathcal{M}}_f$, where $\hat{\mathcal{M}}_u=\mathrm{diag}\left( m_u,\;m_c,\;m_t\right)$  and $\hat{\mathcal{M}}_d=\mathrm{diag}\left( m_d,\;m_s,\;m_b\right)$. 

The singlet leptoquark $S$ introduced in the 331LQ models generate new Yukawa  part  of $S$ with quarks and leptons as follows
\begin{align}
\label{eq:LFVsource}
-\mathcal{L}^{Sff'}=   \sum_{b=1}^3 \left[\sum_{k=1}^2 \widetilde{g}^{LQ}_{kb}\,\overline{Q}^C_{kL}\,L_{bL}S + \sum_{a=1}^3h^{LQ}_{ab}\,\overline{u}^C_{aR}\,e_{bR}S\right] + \mathrm{h.c.}.
\end{align}
 As usual, we denote by $\hat{q}_{uR}$ three right-handed states in the quark mass basis for simplicity. Although only the first two left-handed quark families in the flavor basis couple to $S$, the appearance of the third family's couplings in the physical basis will result in large one-loop contributions to $\Delta a_{e_a}$ \cite{Doff:2024cap}.  The experimental data of the quark mixing matrix $V_{\mathrm{CKM}}\equiv V^{u}_LV^{d\dagger}_L$ must be fixed following the data given in Ref. \cite{ParticleDataGroup:2024cfk}, for example. On the other hand,   the unknown values of leptoquark couplings relating to the third quark family still allow sizable values, as given in the following relations:
 \begin{align}
 \label{eq:hgLQ}
  g^{LQ}_{ib} = -\sum_{k=1}^2 \widetilde{g}^{LQ}_{ki}\,(V_L^{u*})_{ak}, \; g^{\prime\,LQ}_{ib} = \sum_{k=1}^2\widetilde{g}^{LQ}_{kb}\,(V_L^{d*})_{ik}; \; i=1,2,3;\; b=1,2,3. 
 \end{align}
Because none of  $g^{\prime\,LQ}_{ib}$ of the down quarks have  right-handed partners, only one Yukawa part gives sizable one-loop contributions to $(g-2)_{e_a}$ and LFV decay amplitudes of charged leptons, namely \cite{Doff:2024cap}:
\begin{align}
-\mathcal{L}_{Y}^{LQ} =\sum_{i,b=1}^3 \overline{(\hat{u}_{i})^c} \left( g^{LQ}_{ib} P_L + h^{LQ}_{ib} P_R \right) e_b S + \mathrm{h.c.}.
\end{align}
We will pay attention to this LFV source as dominant one-loop contributions to deviations of $(g-2)_{e_a}$ between the 331LQ model and the SM.

\section{\label{sec_coupling} Couplings and analytic formulas for $(g-2)_{e_a}$ anomalies and LFV decay rates}
\begin{figure}[ht]
	\centering 
	\includegraphics[trim=1cm 14.5cm 1cm 3cm, clip, width=1\textwidth]{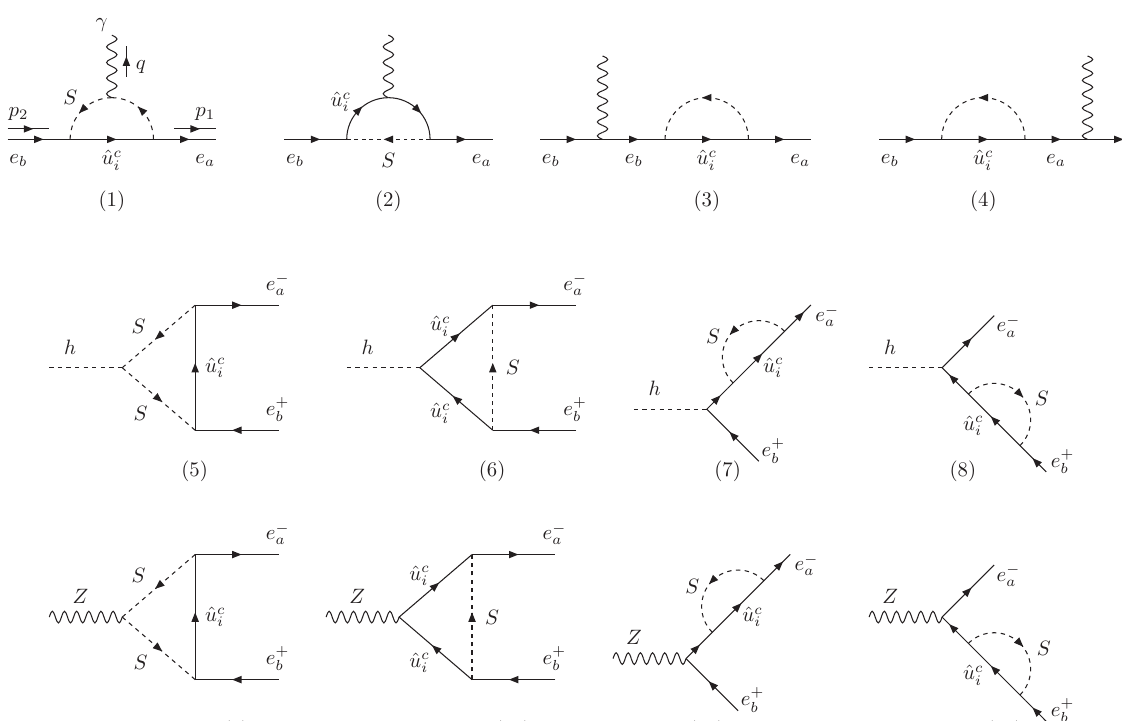}
	\caption{ One-loop Feynman diagrams with leptoquark exchanges contributing to LFV decay amplitudes predicted by the 331LQ framework, where $i=1,2,3$ presents the family index of the up quark. }
	\label{f:1loopLFV}
\end{figure}

We consider here three decay channels cLFV, LFV$h$, and LFV$Z$ decays, in which the decay $\mu\to e\gamma$ is discussed in Ref.~\cite{Doff:2024cap}. Using the general formulas for one-loop contributions to these LFV decay channels introduced in Ref.~\cite{Hue:2024rij}. The relevant one-loop Feynman diagrams in the unitary gauge are depicted in Fig.~\ref{f:1loopLFV},
in which the first, second, and third lines correspond to cLFV, LFV$h$, and LFV$Z$ decay amplitudes. The first two diagrams in the first line were mentioned in Ref. \cite{Doff:2024cap}, while the second lines do not contribute to final formulas of cLFV amplitudes ($a<b$) nor $(g-2)_{e_a}$ ($a=b$) anomalies, see a detailed calculation in Ref. \cite{Hue:2023rks}, for example. 

The analytic formulas of one-loop form factors for leptoquark exchange  relating to $\Delta a_{e_a}$ and cLFV decay amplitudes are \cite{Lavoura:2003xp, Crivellin:2018qmi, Hue:2023rks}:
 \begin{align}
 \label{eq:cabLR}
 c^S_{(ab)R}=& \frac{3e}{16\pi^2m_S^2}\sum_{i=1}^3 \left[m_{u_i} g^{LR}_{ab}  \left(f_S(x_{i})-\frac{2}{3}g_S(x_i)\right)  \right. 
 \crn&\left. \hspace{2.2cm}+ \left(m_{b} g^{LL}_{ab} +m_{a} g^{RR}_{ab}\right) \left( \tilde{f}_S(x_i)-\frac{2}{3} \tilde{g}_S(x_i)\right) \right], 
 \crn  c^S_{(ba)R}=& c_{(ba)R} \left[ g^{LL}_{ab} \leftrightarrow g^{RR}_{ab},  g^{RL}_{ab} \leftrightarrow g^{LR}_{ab}\right],
 \end{align}
 where $x_i=m_{u_i}^2/m_S^2$, the factor $3$ is the color quark index, and  the one-loop master functions  are \cite{Crivellin:2018qmi}
 \begin{align}
 \label{eq:fgS}
 f_S(x)=&2\tilde{g}_S(x)= \frac{x^2 -1- 2x\ln(x)}{4(x-1)^3},
 \crn g_S(x)=&\frac{x-1-\ln(x)}{2(x-1)^2},
 \crn \tilde{f}_S(x)=&\frac{2x^3 +3x^2 -6x +1 -6x^2\ln(x)}{24(x-1)^4}. 
 \end{align}
We also introduce the factor $g^{XY}_{ab}$ with $X,Y=L,R$ are LFV sources derived from Eq. \eqref{eq:LFVsource} as follows
 \begin{align}
 \label{eq:gXLR}
 g^{LL}_{ab} \equiv g^{LQ*}_{ia}g^{LQ}_{ib},\; g^{RR}_{ab} \equiv h^{LQ*}_{ia}h^{LQ}_{ib},\; g^{LR}_{ab} \equiv g^{LQ*}_{ia}h^{LQ}_{ib},\; g^{RL}_{ab} \equiv h^{LQ*}_{ia}g^{LQ}_{ib}. 
 \end{align}
{ We note that although master formulas in Eq. \eqref{eq:fgS} were considered valid for heavy fermions \cite{Crivellin:2018qmi}, we confirm that they are also valid for light fermions, including light quarks discussed in this work.} Respectively, the formulas for $\Delta a_{e_a}$ and cLFV branching ratios (Brs) are:
 \begin{align}
 \label{eq:cLFV}
 \Delta a_{e_a}=& -\frac{4m_{e_a}}{e}\mathrm{Re} \left[c^S_{(aa)R}\right],
 \crn \mathrm{Br}(e_b\to e_a\gamma)=& \frac{48 \pi^2}{G_F^2 m_b^2} \left( \left| c^S_{(ab)R}\right|^2 + \left| c^S_{(ba)R}\right|^2\right) \mathrm{Br}\left( e_b\to e_a \overline{\nu_a}\nu_b\right),
 \end{align}
 where $G_F=1.166378 \times 10^{-5} \mathrm{GeV}$ \cite{ParticleDataGroup:2024cfk}. 
 
 The decay rates of LFV$h$ decays are given by \cite{Pilaftsis:1992st, Arganda:2004bz, Arganda:2014dta}
 \begin{equation}
 \mathrm{Br} (h \rightarrow e_ae_b)\equiv \frac{\Gamma (h\rightarrow e_a^{-} e_b^{+})+\Gamma (h \rightarrow e_a^{+} e_b^{-})}{\Gamma^{\mathrm{total}}_{h}}
 \simeq   \fr{ m_{h}}{8\pi}\left(\vert \Delta^{(ab)}_L\vert^2+\vert \Delta^{(ab)}_R\vert^2\right), \label{eq_LFVwidth}
 \end{equation}
 where $\Gamma^{\mathrm{total}}_{h}\simeq 4.1\times 10^{-3}$ GeV \cite{LHCHiggsCrossSectionWorkingGroup:2016ypw}  and  $\Delta^{(ab)}_{L,R}$ are one-loop contributions arising from leptoquark exchange, namely the one-loop Feynman diagrams are  given in the second line of Fig. \ref{f:1loopLFV}.  The particular analytic formulas are derived from general results shown in Ref. \cite{Hue:2024rij}, namely 
 \begin{align}
  \label{eq:deS}
 \Delta^{(ab)}_{L(R)}=&  \Delta^{(ab)uSS}_{L(R)}+ \Delta^{(ab)Suu}_{L(R)},
 \end{align}
 where $\Delta^{(ab)uSS}_{L(R)}$  is sum of formulas from the three diagrams (5), (7), and (8); while $ \Delta^{(ab)Suu}_{L(R)}$ is from diagram (6).
 \begin{align}
  \Delta^{(ab)uSS}_L=& \frac{3\lambda_{hSS}}{16\pi^2} \sum_{i=1}^3 \left[ {g^{RL}_{ab}}m_{u_i} C_0 -\left(g^{LL}_{ab} m_a C_1 +g^{RR}_{ab}m_b C_2\right) \right]
 \crn &+ \frac{3g}{32 \pi^2m_W(m_a^2-m_b^2)} \sum_{i=1}^3 \left[ g^{LR}_{ab} m_am_b m_{u_i} \left(B^{(1)}_0 -B^{(2)}_0\right) +g^{RL}_{ab} m_{u_i} \left(m_b^2 B^{(1)}_0 -m_a^2 B^{(2)}_0\right)
 \right. \crn & \hspace{4cm}-\left. m_am_b \left(g^{LL}_{ab} m_b + g^{RR}_{ab} m_a \right)  \left(B^{(1)}_1 -B^{(2)}_1\right)\right],
 \crn 	\Delta^{uSS}_R= &	\Delta^{uSS}_L\left[ g^{LL}_{ab} \leftrightarrow g^{RR}_{ab},  g^{RL}_{ab} \leftrightarrow g^{LR}_{ab}\right],
 \end{align}
 where  $\Delta^{(ab)uSS}_{L(R)}$ in Eq. \eqref{eq:deS} are written in terms of  well-known  Passarino-Veltman functions \cite{Passarino:1978jh}:  $C_{i}=C_{i}(m_a^2,m_h^2,m_b^2; m_{u_i}^2,m_S^2,m_S^2)$ with $i=0,1,2$;  and $B^{(k)}_{0,1}=B^{(k)}_{0,1}(p_k^2;m^2_{u_i},m^2_S)$ ($k=1,2$), using notations defined precisely in Ref. \cite{Hue:2024rij}, based on LoopTools \cite{Hahn:1998yk} implemented in our numerical investigation. The coupling factor $\lambda_{hSS}$ was derived from the Higgs potential given in Appendix \ref{app:higgs}. The Yukawa  factors relating to couplings of the SM-like Higgs boson with two anti-up quarks $g^{L(R)}_{\hat{u}^c_i\hat{u}^c_i}$ derived  by identifying from the general part $\mathcal{L}^Y=-h\sum_{i=1}^3\hat{u}^c_{i} \left[ g^{L}_{\hat{u}^c_i\hat{u}^c_i} P_L + g^{R}_{\hat{u}^c_i\hat{u}^c_i} P_R\right]\hat{u}^c_{i} +\mathrm{h.c.}$   with Lagrangian \eqref{Lmass}. In particular,  the Lagrangian with physical states are:
  \begin{align}
  \label{eq:Lyhqq}
  -\mathcal{L}^Y_{hqq}=&g^d_{ka}\overline{d}_{kL}\eta^{0*}_1d_{aR} + g^d_{3a}\overline{d}_{3L}\rho^{0} d_{aR}  - g^u_{ka}\overline{u}_{kL}\rho^{0}u_{aR} + g^u_{3a}\overline{u}_{3L}\eta^{0*}_1 u_{aR} +\mathrm{h.c.}
  \crn =&g^d_{ka} \frac{s_{\beta}h+vs_{\beta}}{\sqrt{2}}\overline{d}_{kL}d_{aR} + g^d_{3a} \frac{c_{\beta}h+vc_{\beta}}{\sqrt{2}}\overline{d}_{3L} d_{aR}  
  \crn&-  g^u_{ka}\frac{c_{\beta}h+vc_{\beta}}{\sqrt{2}}\overline{u}_{kL}u_{aR} + g^u_{3a} \frac{s_{\beta}h+vs_{\beta}}{\sqrt{2}} \overline{u}_{3L} u_{aR} +\mathrm{h.c.} +\dots 
  \crn =& \left( 1+\frac{h}{v}\right) \sum_{f=u,d}\overline{q_{fL}}\mathcal{M}_fq_{fR} +\mathrm{h.c.} +\dots 
  \crn =&\left( 1+\frac{h}{v}\right) \sum_{f=u,d}\overline{\hat{q}_{fL}}\hat{\mathcal{M}}_f\hat{q}_{fR} +\mathrm{h.c.} +\dots 
  \end{align}
 The result in Eq. \eqref{eq:Lyhqq} shows that couplings of $h$ with SM-like quarks are exactly the same as those from SM, therefore $g^{L}_{\hat{u}^c_i\hat{u}^c_i}=g^{R}_{\hat{u}^c_i\hat{u}^c_i}=gm_{u_i}/(2m_W)$. As a result, $\Delta^{(ab) Suu}_{L(R)}$ has the following simple formulas:
   \begin{align}
  \label{eq:DeSuu}
    \Delta^{(ab) Suu}_L=&\frac{3g}{ 32\pi^2 m_W}\sum_{i=1}^3   \left\{   m_{u_i} \left[  g^{RL}_{ab} \left( B^{(12)}_0 +(m_{u_i}^2+ m_S^2) C_0 +m_a^2 C_1 +m_b^2 C_2 \right) +g^{LR}_{ab} m_am_bX_0
  \frac{}{}\right.\right. \crn& \left. \left. \frac{}{}\hspace{3.2cm} + g^{LL}_{ab} m_a m_{u_i} (C_0 +2C_1) +g^{RR}_{ab} m_b m_{u_i} (C_0 +2C_2)  \right]\right\},
  \crn 	\Delta^{Suu}_R= &	\Delta^{Suu}_L\left[ g^{LL}_{ab} \leftrightarrow g^{RR}_{ab},  g^{RL}_{ab} \leftrightarrow g^{LR}_{ab} \right],
  \end{align}
  where $g^{XY}_{ab}$ are given in Eq. \eqref{eq:gXLR}, and the PV-functions are  $C_{i}=C_{i}(m_a^2,m_h^2,m_b^2; m_S^2,m_{u_i}^2, m_{u_i}^2)$ with $i=0,1,2$; and $B^{(12)}_{0}=B_{0,1}(m_h^2;m^2_{u_i},m^2_{u_i})$.  We can see that although both $\Delta^{uSS}_{L(R)}$ and $\Delta^{Suu}_{L(R)}$ contains divergences, the final sum of them satisfies the property of  divergent cancellation. In particular, from the property of PV-functions \cite{Hue:2024rij}, get:
  $$ \mathrm{div} \left[ \Delta^{uSS}_{L} \right] + \mathrm{div} \left[ {\Delta^{Suu}_{L}} \right]\varpropto \mathrm{div}\left[B^{(12)}_0\right] +\frac{m_b^2 \mathrm{div}\left[B^{(1)}_0\right] -m_a^2 \mathrm{div}\left[B^{(2)}_0\right]}{m_a^2-m_b^2} =0. $$

 The decay rates of LFV$Z$ decays are given by $\mathrm{Br}(Z\to e^+_b e^-_a)= \Gamma (Z\to e^+_b e^-_a)/\Gamma_Z$, where the total decay width of the $Z$ boson is $\Gamma_Z= 2.4955$ GeV  \cite{ParticleDataGroup:2024cfk}, and \cite{Korner:1992an, DeRomeri:2016gum, Jurciukonis:2021izn, Hong:2023rhg}:
 \begin{align} 
 \label{eq_GAZeba}
 \Gamma (Z\to e^+_b e^-_a)= 	\frac{\sqrt{\lambda}}{16\pi m_Z^3}\times \left(\frac{e}{16\pi^2}\right)^2 \left( \frac{\lambda M_0}{12 m^2_Z} +M_1 +\frac{ M_2}{3 m^2_Z}\right),
 \end{align}
 where $\lambda= m^4_Z +m^4_{b} +m^4_{a} -2(m^2_Zm^2_{a} +m^2_Zm^2_{b} +m^2_{a}m^2_{b})$, 
 and the formulas of $M_{0,1,2}$ were given in Ref. \cite{Jurciukonis:2021izn}, which are presented in a reduced form \cite{Hong:2023rhg}
 \begin{align}
 \label{eq_Mi}
 M_0= & (m^2_Z -m_{a}^2 -m_{b}^2)\left(|\bar{b}^{331\mathrm{LQ}}_L|^2 +|\bar{b}^{331\mathrm{LQ}}_R|^2\right)  -4 m_{a} m_{b} \mathrm{Re}\left[ \bar{b}^{331\mathrm{LQ}}_L  \bar{b}^{331\mathrm{LQ}*}_R\right]
 \crn&
 - 4m_{b} \mathrm{Re}\left[ \bar{a}^{331\mathrm{LQ}*}_R \bar{b}^{331\mathrm{LQ}}_L   + \bar{a}^{331\mathrm{LQ}*}_L \bar{b}^{331\mathrm{LQ}}_R  \right] -  4m_{a}\mathrm{Re}\left[ \bar{a}^{331\mathrm{LQ}*}_L \bar{b}^{331\mathrm{LQ}}_L   + \bar{a}^{331\mathrm{LQ}*}_R  \bar{b}^{331\mathrm{LQ}}_R  \right] , 
 \crn M_1 = & 4 m_{a}m_{b} \mathrm{Re}\left[\bar{a}^{331\mathrm{LQ}}_L\bar{a}^{331\mathrm{LQ}*}_R \right],
 \crn  M_2 = &  \left[ 2 m^4_Z - m_Z^2\left( m_{a}^2 + m_{b}^2\right) - \left( m_{a}^2 - m_{b}^2\right)^2  \right] \left( |\bar{a}^{331\mathrm{LQ}}_L|^2 +|\bar{a}^{331\mathrm{LQ}}_R|^2\right).
 \end{align}
 where we omit the LFV index $(ab)$ in the right handed side for simplicity.  
 
 The contributions from diagrams with pure scalar exchanges were shown previously in Ref. \cite{,Hue:2024rij}.  Particular formulas of the amplitudes are written as follows.  Final results for form factors corresponding to diagram (5) in Fig. \ref{f:1loopLFV} are
 \begin{align}
 \label{eq:abLR}
 \overline{a}^{331\mathrm{LQ}}_{L(R)}= \bar{a}^{uSS}_{L(R)}+\bar{a}^{Suu}_{L(R)},\;  \overline{b}^{331\mathrm{LQ}}_{L(R)}= \bar{b}^{uSS}_{L(R)}+\bar{b}^{Suu}_{L(R)},
 \end{align}
  where sum of one-loop contributions from three diagrams (9), (11), and (12) result in the following form factors  
 \begin{align}
 \label{eq_ab4LR}	
 \bar{a}^{uSS}_{L}&=- 6g_{ZSS} \sum_{i=1}^3 g^{LL}_{ab} C_{00}
 \crn  &\quad - \frac{ 3t_{L}}{m_a^2 -m_b^2}\sum_{i=1}^3  \left[  m_{u_i} \left( m_a g^{RL}_{ab}  + m_b g^{LR}_{ab}   \right) \left(B^{(1)}_0 -B^{(2)}_0\right) 
 \right. \crn& \left. \hspace{3.4cm} -m_am_b g^{RR}_{ab}\left( B^{(1)}_1-B^{(2)}_1\right)   - g^{LL}_{ab}   \left(m_a^2B^{(1)}_1 -m_b^2 B^{(2)}_1 \right) 
 \right],
 \crn  \bar{b}^{uSS}_{L}& = -2g_{ZSS} \sum_{i=1}^3 \left[   m_a  g^{LL}_{ab} X_1  +  m_b  g^{RR}_{ab} X_2  -m_{u_i} g^{RL}_{ab}  X_0  \right] ,
 \crn  \bar{a}^{uSS}_{R}& = \bar{a}^{uSS}_{L} \left[ t_L\to t_R, g^{LL} \leftrightarrow g^{RR}, g^{RL} \leftrightarrow g^{LR} \right],
 \crn  \bar{b}^{uSS}_{R}& =\bar{b}^{uSS}_{L} \left[ g^{LL} \leftrightarrow g^{RR}, g^{RL} \leftrightarrow g^{LR} \right],
 \end{align}
 where  $g^{XY}$ is given in Eq. \eqref{eq:gXLR}, and  arguments of the PV-funtions are $(m_{u_i}^2, m_{S}^2, m_{S}^2)$,  $B^{(k)}_{0,1}=B_{0,1}(p_k^2;m_{u_i}^2,m_S^2)$ with $k=1,2$. The factor coupling $g_{ZSS}$ is derived from the kinetic terms
  $$\mathcal{L}_{\mathrm{kin}}^S=(D_{\mu}S)^*(D^{\mu}S) =-ieg_{ZSS}Z_{\mu}SS^*(p_{S^*}-p_S)^{\mu}+\dots,$$
where  $D_{\mu}S=\left( \partial_{\mu}-i \frac{g_XX_S}{\sqrt{6}}X_{\mu} \right) S$. Using $g_X=3\sqrt{2}g{s_W}/\sqrt{3-4s_W^2}$ and $X_{\mu}\to -s_W \sqrt{1-t_W^2/3} Z_{\mu}$, as well-known in the 3-3-1 model with neutral leptons \cite{Long:1995ctv,Hong:2024swk}, and $X_S=1/3$, we get $g_{ZSS}={t_W/3}$.
  
 Form factors corresponding to diagram (10) are
 \begin{align}
 \label{eq_ab6LR}	
 \bar{a}^{Suu}_{L}=& -3 \sum_{i=1}^3 \left\{  g^{L}_{Z\hat{u}^c_i \hat{u}^c_i}\left[ g^{LL}_{ab}  m_{u_i}^2 C_0 + g^{RL}_{ab}  m_{a} m_{u_i}(C_0+C_1) \frac{}{}
 \right.\right.\crn&\left.\left.\qquad\qquad\qquad\frac{}{}
 +  g^{LR}_{ab}  m_{b} m_{u_i}(C_0+C_2) +  g^{RR}_{ab} m_{a} m_bX_0 \frac{}{}\right]   
 \right.\crn  &\left. \qquad \quad-g^{R}_{Z\hat{u}^c_i \hat{u}^c_i} \left[ g^{LL}_{ab} \left( (d-2)C_{00} +m_a^2 X_1 +m_b^2X_2 -m_Z^2 C_{12}\right) \frac{}{}
 \right.\right.\crn&\left.\left.\qquad \qquad \qquad \; \frac{}{} +m_am_{u_i}g^{RL}_{ab}C_1 + m_bm_{u_i}g^{LR}_{ab}C_2 \right]\right\}
 \crn \bar{b}^{Suu}_{L}=& -6 \sum_{i=1}^3 \left[\frac{}{} g^{L}_{Z\hat{u}^c_i \hat{u}^c_i}  \left( g^{RL}_{ab} m_{u_i}C_2 +  g^{RR}_{ab}  m_{b} X_2\right) 
 %
 +g^{R}_{Z\hat{u}^c_i \hat{u}^c_i}  \left(  g^{RL}_{ab}  m_{u_i} C_1 +  g^{LL}_{ab}  m_{a} X_1\right) 
 \right],
 \crn  \bar{a}^{Suu}_{R}=& \bar{a}^{Suu}_{L} \left[ g^{L}_{Z\hat{u}^c_i \hat{u}^c_i} \leftrightarrow g^{R}_{Z\hat{u}^c_i \hat{u}^c_i}, g^{LL} \leftrightarrow g^{RR}, g^{RL} \leftrightarrow g^{LR}\right],
 \crn\bar{b}^{S uu}_{R} =& \bar{b}^{Suu}_{L} \left[ g^{L}_{Z\hat{u}^c_i \hat{u}^c_i} \leftrightarrow g^{R}_{Z\hat{u}^c_i \hat{u}^c_i}, g^{LL} \leftrightarrow g^{RR}, g^{RL} \leftrightarrow g^{LR} \right], 
 \end{align}
 where $g^{XY}$ is given in Eq. \eqref{eq:gXLR} and   arguments of the PV-functions are $(m_S^2,m^2_{{u}_i},m^2_{{u}_i})$.  The coupling factors $g^{L,R}_{Z\hat{u}^c_i\hat{u}^c_i}$ derived from the general form $\mathcal{L}^{Zff}=e Z^{\mu}\sum_{q}\overline{q} \gamma_{\mu}\left[g^{L}_{Zqq} P_L + g^{R}_{Zqq} P_R\right]q +\mathrm{h.c.}$. For the particular case of the 331LQ model, see for example the detailed formulas in Ref. \cite{Hung:2019jue}. We consider here the limit of very large {$v_\chi$} scale so the $Zqq$ couplings are exactly the same as those in the SM, as listed in Table \ref{t:Zff} for the up-type antiquarks and charged leptons.
 \begin{table}[h]
 	\begin{tabular}{|c|c|c|}
 		\hline
 	f& $g^L_{Zff}$ & 	$g^R_{Zff}$  \\
 		\hline
 		$e_a$&$\frac{2s_W^2-1}{2s_Wc_W}=t_L$& $t_W=t_R$\\
  \hline 
  	$\hat{u}^c_i=u^c,c^c,t^c$&$\frac{2 t_W}{3}$& $\frac{-1}{s_Wc_W}\left(\frac{1}{2} -\frac{2}{3}s_W^2\right)$\\
  \hline 
 	\end{tabular}
 	\caption{Coupling factors for the $Z$ boson interactions with two fermions.
 		\label{t:Zff}}
 \end{table}
 The Feynman rules involving up-type antiquarks are obtained from those of the corresponding quarks by using the identity $\overline{\hat{u}^c_i} \gamma^{\mu} P_{L(R)}\hat{u}^c_i=- \overline{\hat{u}_i} \gamma^{\mu} P_{R(L)}\hat{u}_i$ \cite{Dreiner:2008tw}.

In numerical investigations, we comment here on interesting qualitative properties of analytic formulas for the one-loop contributions to $\Delta a_{e_a}$ and LFV decay amplitudes. We will focus on the most interesting regions predicting at least one of the  following: $|\Delta a_{\mu}|\geq 10^{-11}$ or/and  $ |10^{-14}| \leq |\Delta a_{e}| \leq 5\times 10^{-13}$. In addition, all current experimental constraints on AMMs and LFV searches must be satisfied.  Since the dominant contributions to the mentioned quantities are proportional to the quark masses, those induced by the two light up-type quarks are neglected. This feature will be confirmed numerically with the following relations of AMMs and cLFV decays:
\begin{align}
\label{eq:LFVreduced} 
&\Delta a_{e_a}\varpropto \mathrm{Re}[m_t g^{LQ*}_{3a} h^{LQ}_{3a}],
\crn& \mathrm{Br}(e_b\to e_a\gamma)\;  \varpropto f_{ba}\equiv  \left( \left| g^{LQ*}_{3a} h^{LQ}_{3b}\right|^2 + \left| h^{LQ*}_{3a} g^{LQ}_{3b}\right|^2 \right).
\end{align}
We will also check the interesting fact that $\mathrm{Br}(h,Z\to e_b e_a)  \varpropto f_{ba} \propto \mathrm{Br}(e_b\to e_a\gamma)$ as consequences can be seen directly from analytic formulas of LFV decay amplitudes.

\section{\label{sec_numerical} Numerical discussion}
In this section, we will use the experimentally determined parameters are \cite{ParticleDataGroup:2024cfk}: $g  =0.652, G_F=1.166378 \times 10^{-5} \mathrm{GeV^{-2}}, s_W^2=0.231, m_W=80.3692 \mathrm{GeV}$, $	m_e  =5 \times 10^{-4} \mathrm{GeV}, m_\mu=0.105 \mathrm{GeV}, m_\tau=1.777 \mathrm{GeV}, m_Z=91.1880 \mathrm{GeV}, m_t = 172,57 \mathrm{GeV}$. The scanning ranges of free parameters are chosen as follows:
\begin{align}
\label{eq:scanning}
	&m_{S}  \in[0.5,5]\, \mathrm{TeV}, \lambda_{hSS} \in[-5,5] \times v_0,\, |g(h)^{LQ}_{3i}|\leq \sqrt{4\pi}\, \forall i=1,2,3,
\end{align}
where the couplings $g(h)^{LQ}_{3i}$ alway satisfy the perturbative limit. We note here that although the couplings  $g^{LQ}_{3i}$ depends the experimental data of  the quark mixing matrix $V_{\mathrm{CKM}}$,  as given in Eq. \eqref{eq:hgLQ}, the unknown property of  $V^u_L$ ($V^d_L$) still allows large values of $g^{LQ}_{3i}$ close to the perturbative bounds.

\begin{figure}[ht]
	\centering
	\begin{tabular}{cc}
		\includegraphics[width=8.5cm]{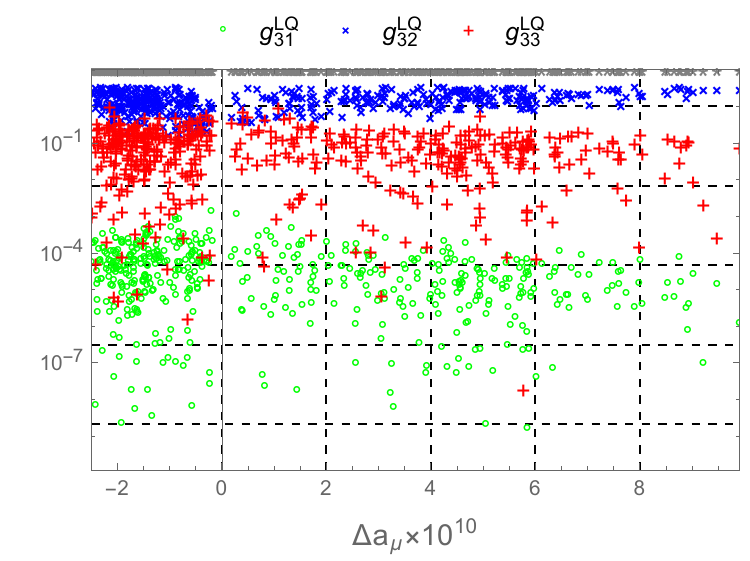} &
		\includegraphics[width=8.5cm]{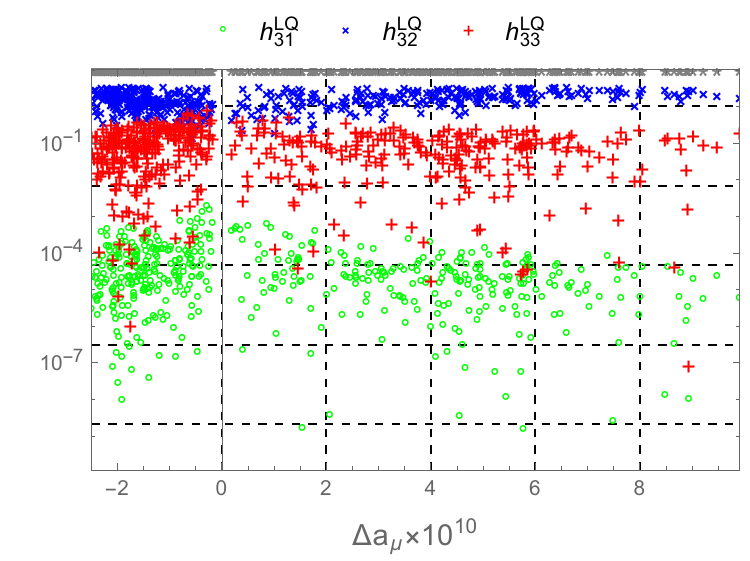} \\
		\includegraphics[width=8.5cm]{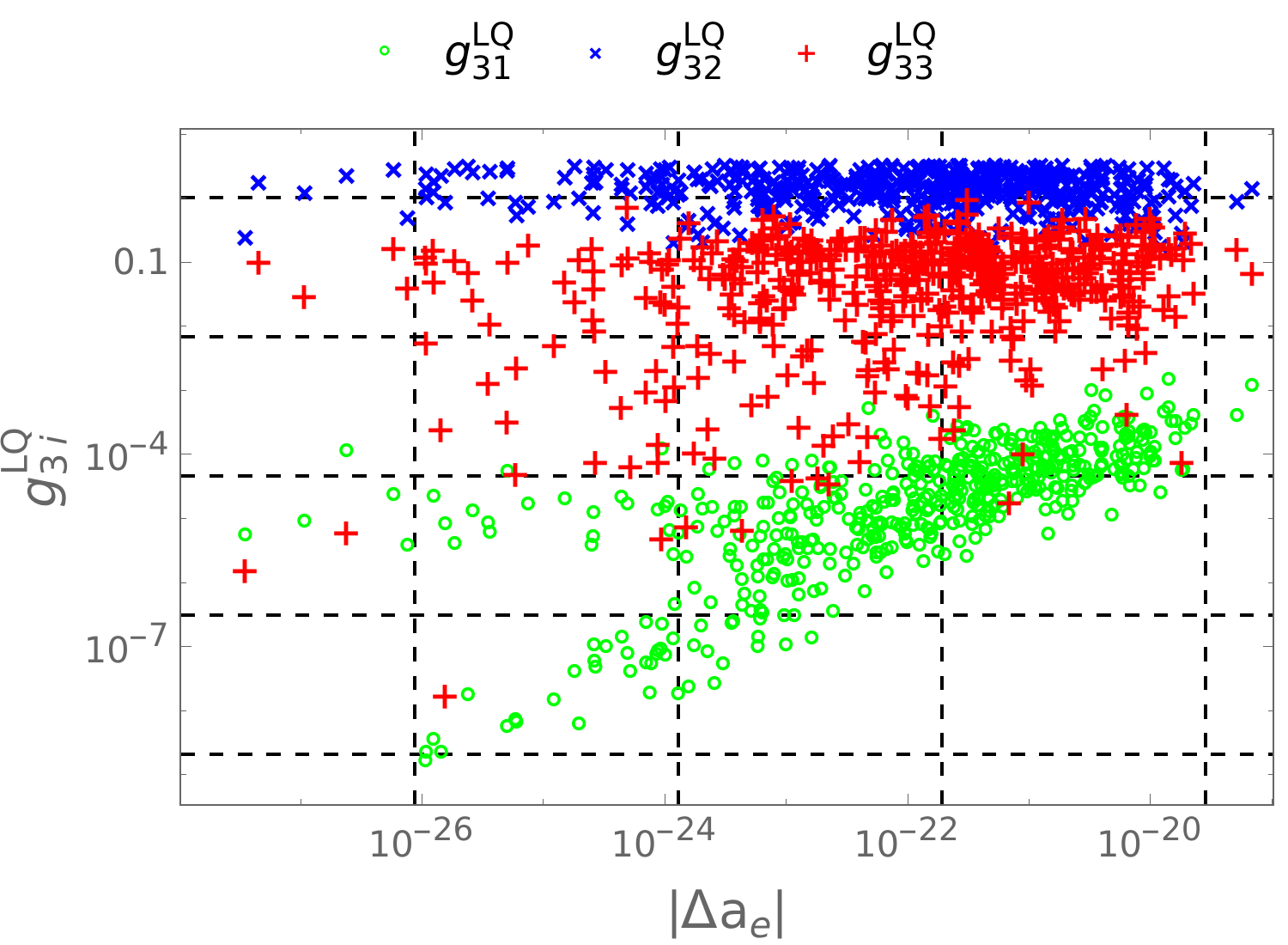} &
		\includegraphics[width=8.5cm]{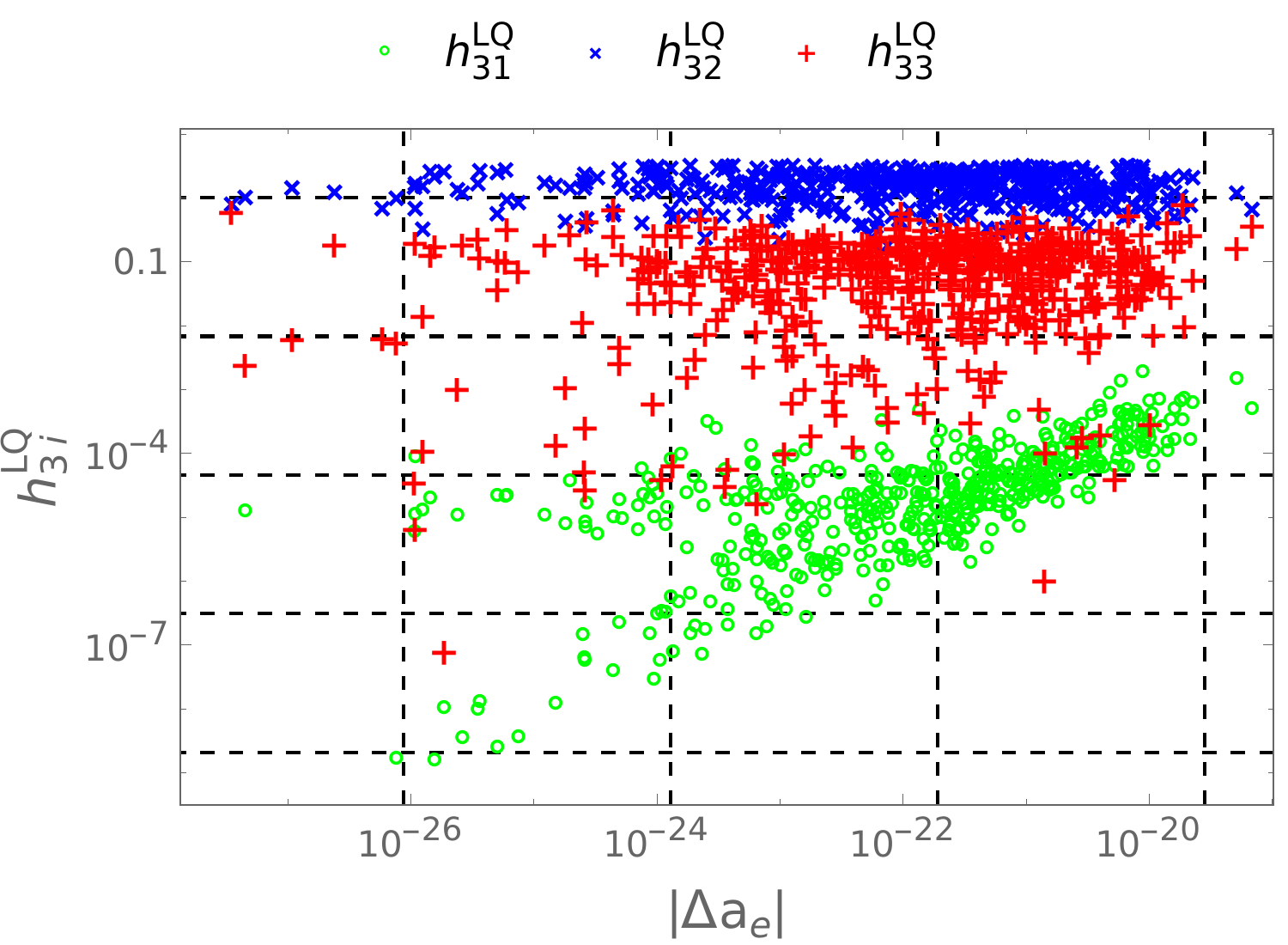} \\
	\end{tabular}
	\caption{The relationship between the AMMs of muon $\Delta {a_\mu}$ and  electron $|\Delta {a_e}|$  with respect to the Yukawa couplings $g(h)_{3i}^{LQ}$ shown in left(right)-panel, respectively.}\label{fig_aegh}
\end{figure}

The numerical results presented here  are allowed regions of parameter space consisting of allowed points that satisfy all current experimental constraints of  AMMs and LFV decay rates that will be shown precisely, including the deviation from muon AMM given in Eq. \eqref{eq:damu}. We start our numerical discussion by investigating the relevance of the electron and muon AMMs $\Delta a_{e,\mu}$ on the leptoquark Yukawa couplings $g_{3i}^{\mathrm{LQ}}$ and $h_{3i}^{\mathrm{LQ}}$, as shown in Fig.~\ref{fig_aegh}.
It is evident that $|\Delta a_e|$ exhibits no noticeable dependence on $g_{32,33}^{\mathrm{LQ}}$ or $h_{32,33}^{\mathrm{LQ}}$, whereas it increases significantly with increasing $g_{31}^{\mathrm{LQ}}$ and $h_{31}^{\mathrm{LQ}}$, which can be seen precisely from Eq.~ \eqref{eq:cabLR}, where $g^{LR}_{12}=g^{LQ*}_{31}h^{LQ}_{31}$ for top-quark derived from Eq.~\eqref{eq:gXLR}. Similarly, sizable values of  $\Delta a_{\mu}$ support only large $g(h)^{LQ}_{32}$. This behavior is due to the fact that the dominant one-loop contribution to $a_e$ is governed by the couplings involving the first-generation charged lepton. {Additionally, the predicted values are below $10^{-19}$, which is much smaller than the  current discrepancy of  $\mathcal{O}(10^{-13})$  between experimental measurements reported by different groups  \cite{Hanneke:2008tm, Parker:2018vye, Morel:2020dww, Fan:2022eto} and the SM  prediction.}

\begin{figure}[ht]
	\centering
	\begin{tabular}{cc}
		\includegraphics[width=8.cm]{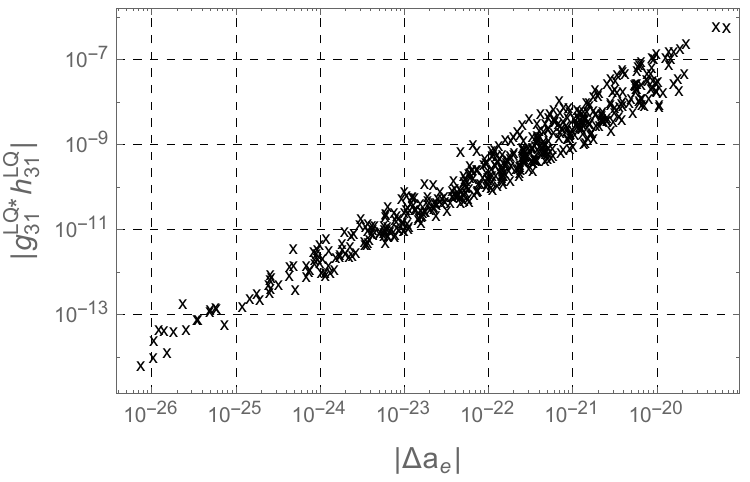} &
		\includegraphics[width=8.cm]{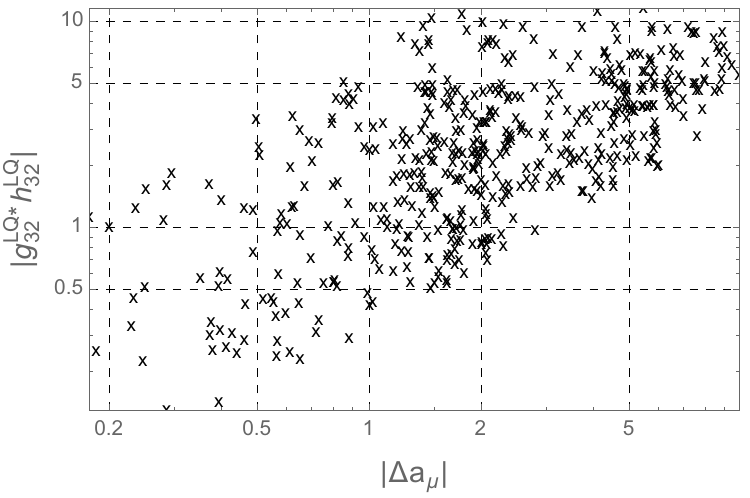} \\
		\includegraphics[width=8cm]{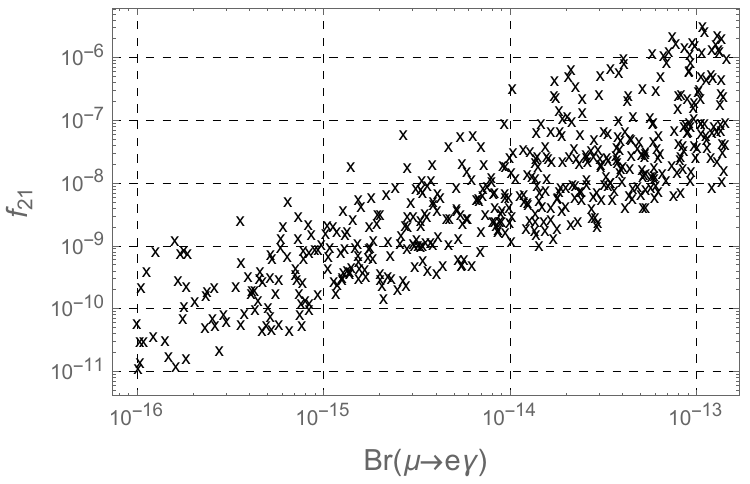} &
		\includegraphics[width=8cm]{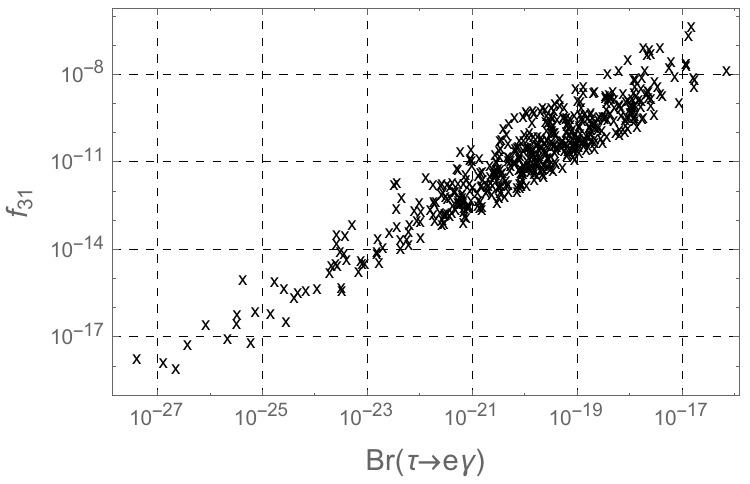} \\
	\end{tabular}
	\caption{The relationship between the AMMs $|\Delta a_{e,\mu}|$ and cLFV decay rates  with respect to combinations of the Yukawa couplings $g(h)_{3i}^{LQ}$ shown in Eq. \eqref{eq:LFVreduced}.}\label{fig_fgh}
\end{figure}

\begin{figure}[ht]
	\centering
	\includegraphics[width=8.5cm]{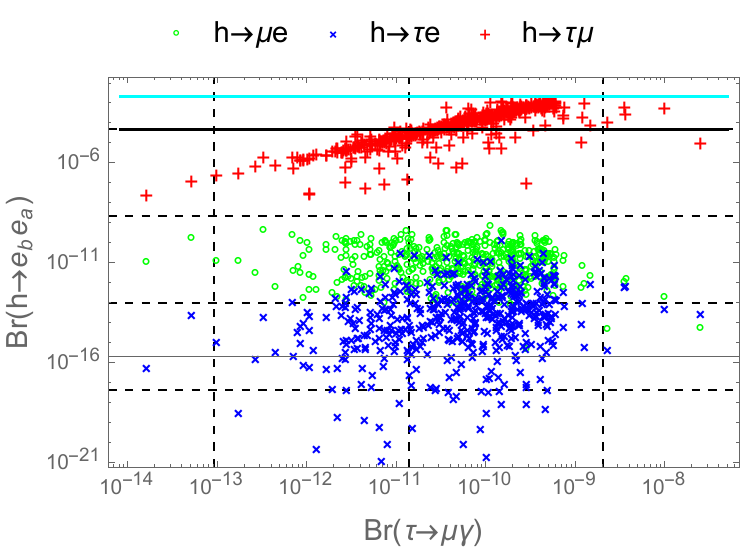}
\begin{tabular}{cc}
	\includegraphics[width=8cm]{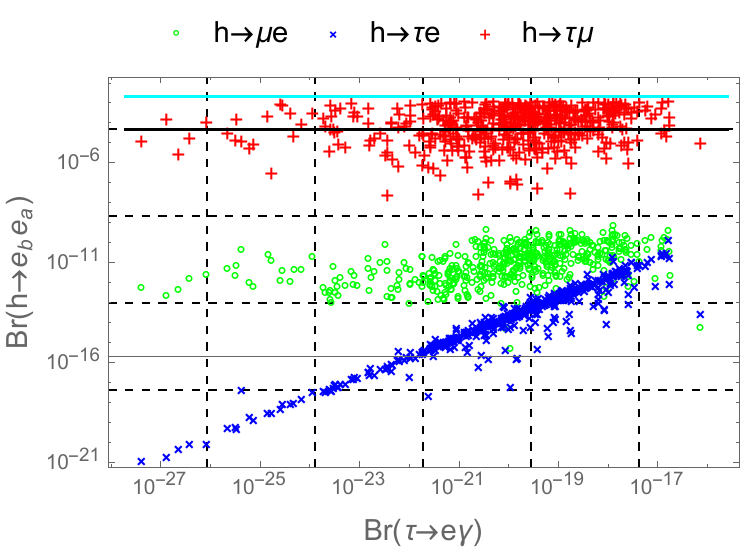} &
	\includegraphics[width=8cm]{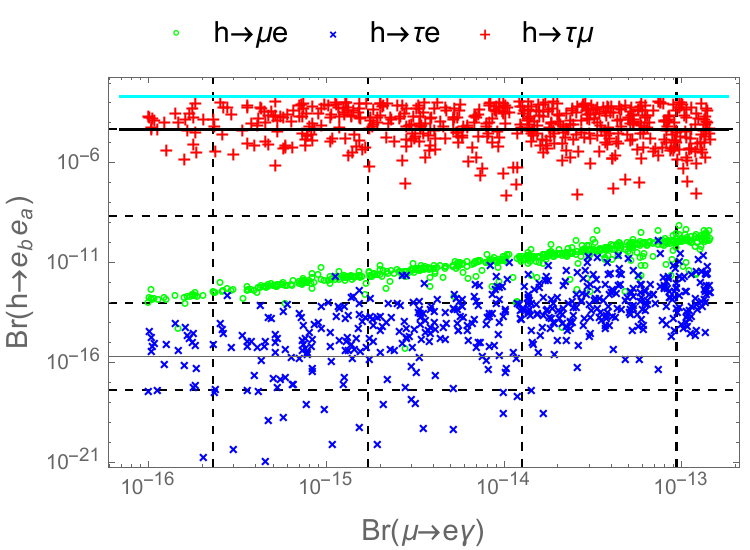} 
\end{tabular}
	\caption{The relationship between decay rates of LFV$h$ with respect to decay rates of cLFV. In each panel, the two cyan and black horizontal lines denote the current experimental upper bounds in Eq.~\eqref{LFV_exp}. The corresponding values are $2.0\times 10^{-3}$ for $\text{Br}(h\to\tau e)$ and $4.4\times 10^{-5}$ for $\text{Br}(h\to\mu e)$, respectively.}\label{fig_LFVheba}
\end{figure}

The above properties of $\Delta a_{e,\mu}$ are shown precisely in Fig. \ref{fig_fgh},
 which  originate from the following: i) dominant contributions given in Eq. \eqref{eq:LFVreduced}; ii) $|10^{-11}|\leq |\Delta a_{\mu}|$ and $-2\times 10^{-10}\leq \Delta a_{\mu} \leq 10^{-9}$; iii) experimental constraint on Br$(\mu \to e\gamma) <1.5 \times 10^{-13}$. The numerical results shown in Fig. \ref{fig_fgh} confirm the dominant parts in  Eq. \eqref{eq:LFVreduced}  over the total analytical formulas of one-loop contributions to $\Delta a_{e_a}$ and cLFV amplitudes. Now, sizable $|\Delta a_{\mu}|$ requires sizable $|g^{LQ}_{32}|$ and $|h^{LQ}_{32}|$, while satisfying the perturbative constraints. On the other hand, small experimental upper bound of Br$(\mu \to e\gamma)$ requires both small values of  $|g^{LQ*}_{31}h^{LQ}_{32}|$ and $|g^{LQ*}_{32}h^{LQ}_{31}|$, therefore result in that both $|g^{LQ}_{31}|$ and $|h^{LQ}_{31}|$ must be tiny. This explains why $|\Delta a_e|$ and  Br$(\tau \to e\gamma)$ are much more suppressed than current experimental sensitivities:  $|\Delta a_e| <10^{-19}$ and Br$(\tau \to e\gamma)<10^{-16}$. Our numerical results of $g(h)^{LQ}_{31,32}$ from constraints of Br$(\mu \to e\gamma)$ and $|\Delta a_{\mu}|$  agree with Ref. \cite{Doff:2024cap}. In addition, the current (expected) experimental bounds \cite{Venturini:2024keu, ParticleDataGroup:2024cfk, MEGII:2025gzr, MEGII:2018kmf, Belle-II:2018jsg,CMS:2023pte,Belle:2021ysv,Qin:2017aju,CMS:2021rsq,Barman:2022iwj,Aoki:2023wfb,ATLAS:2022uhq,ATLAS:2021bdj,ATLAS:2023mvd, Dam:2018rfz, FCC:2018byv} for LFV decay rates are as follows:
\begin{align}\label{LFV_exp}
&-\, \mathrm{The\, cLFV\, decays\,}: \mathrm{Br}(\mu\rightarrow e\gamma) < 1.5\times 10^{-13} \left(<6\times 10^{-14}\right),\crn
& \mathrm{Br}(\tau\rightarrow \mu\gamma) <4.2\times 10^{-8}\left(< 6.9 \times 10^{-9}\right),\,\mathrm{Br}(\tau\rightarrow e\gamma) <3.3\times 10^{-8}\left(< 9.0 \times 10^{-9}\right);
\crn
&-\, \mathrm{The\, LFV\mathit{h}\, decays\,}:  \mathrm{Br}(h\rightarrow \mu e)<4.4\times 10^{-5}\left(\sim\mathcal{O}(10^{-5})\right),\crn
&\mathrm{Br}(h\rightarrow \tau\mu)<1.5\times 10^{-3}\left(\sim\mathcal{O}(10^{-4})\right),\, \mathrm{Br}(h\rightarrow \tau e)<2.0\times 10^{-3}\left(\sim\mathcal{O}(10^{-4})\right);
\crn
&-\, \mathrm{The\, LFV\mathit{Z}\, decays\,}: \mathrm{Br}(Z\to \mu^\pm e^\mp) \leq 2.62\times 10^{-7} \left(10^{-10} \right),\crn
& \mathrm{Br}(Z\to \tau^\pm e^\mp) \leq 5.0\times 10^{-6} \left(10^{-9}\right),\, \mathrm{Br}(Z\to \tau^\pm\mu^\mp) \leq 6.5\times 10^{-5} \left(10^{-9}\right).
\end{align}
Secondly, Fig.~\ref{fig_LFVheba} presents the correlations between the LFV$h$ decay branching ratios and their corresponding cLFV radiative decays. 
A notable feature is the strong correlation observed between each LFV$h$ decay channel and its corresponding cLFV process. As illustrated in the top panel, $\mathrm{Br}(h\to\tau\mu)$ increases monotonically with $\mathrm{Br}(\tau\to\mu\gamma)$, while exhibiting no significant dependence on the other two cLFV channels. Likewise, the left- and right-panels in the bottom show that $\mathrm{Br}(h\to\tau e)$ and $\mathrm{Br(}h\to\mu e)$ are strongly correlated with $\mathrm{Br}(\tau\to e\gamma)$ and $\mathrm{Br}(\mu\to e\gamma)$, respectively. It is worth emphasizing that the present experimental upper bounds on $\mathrm{Br}(\tau\to\mu\gamma)$ and $\mathrm{Br}(\mu\to e\gamma)$, namely $4.2\times10^{-8}$ \cite{ParticleDataGroup:2024cfk, Belle:2021ysv} and $1.5\times10^{-13}$ \cite{Venturini:2024keu, ParticleDataGroup:2024cfk, MEGII:2025gzr}, respectively, still allow $\mathrm{Br}(h\to\tau\mu)$ to reach the current experimental limit of $1.5\times10^{-3}$~\cite{CMS:2021rsq,ParticleDataGroup:2024cfk}. In contrast, the predicted branching ratios $\mathrm{Br}(h\to\tau e)$ and $\mathrm{Br}h\to\mu e)$ remain several orders of magnitude below their corresponding experimental upper bounds. In conclusion, the strong correlations of $\mathrm{Br}(e_b\to e_a\gamma)$ and Br$(h\to e_be_a)$ can be explained easily based on the properties given in Eq. \eqref{eq:LFVreduced} that both of them are mainly proportional to $f_{ba}$. Therefore, in the regions of the parameter space predicting $|\Delta a_{\mu}|\geq 10^{-11}$, the stringent constraint of $\mathrm{Br}(\mu \to e\gamma)$ will result in supressed values of the following LFV$h$ decay rates: $\mathrm{Br}(h\to \mu e)$, $\mathrm{Br}(h\to \tau e)<10^{-9}$. Consequently, future improvements in the experimental sensitivities to the cLFV radiative decays will further restrict the allowed parameter space and, in turn, lead to more stringent predictions for the corresponding LFV$h$ decays. {It is also interesting to note that the upper bounds of these LFV decay channels are in good agreement with those obtained in the recently proposed 3-4-1 model with inverse seesaw neutrinos~\cite{Nha:2026pvq}.}

\begin{figure}[ht]
	\centering
\begin{tabular}{cc}
	\includegraphics[width=8cm]{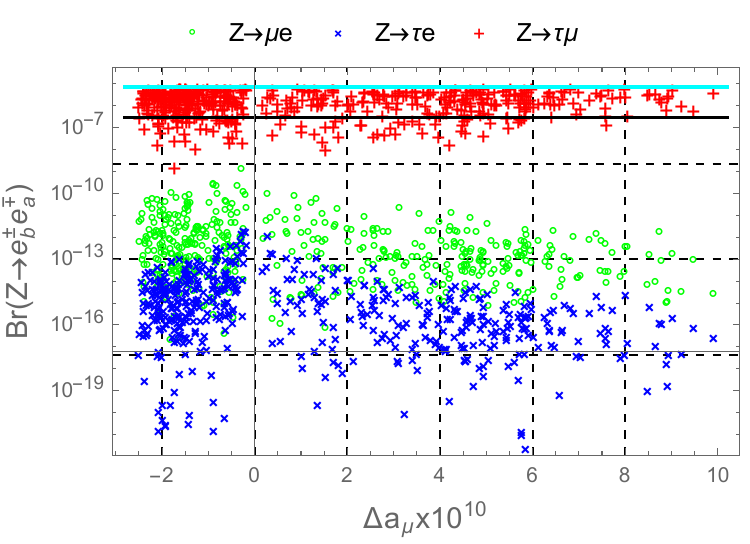} &
	\includegraphics[width=8cm]{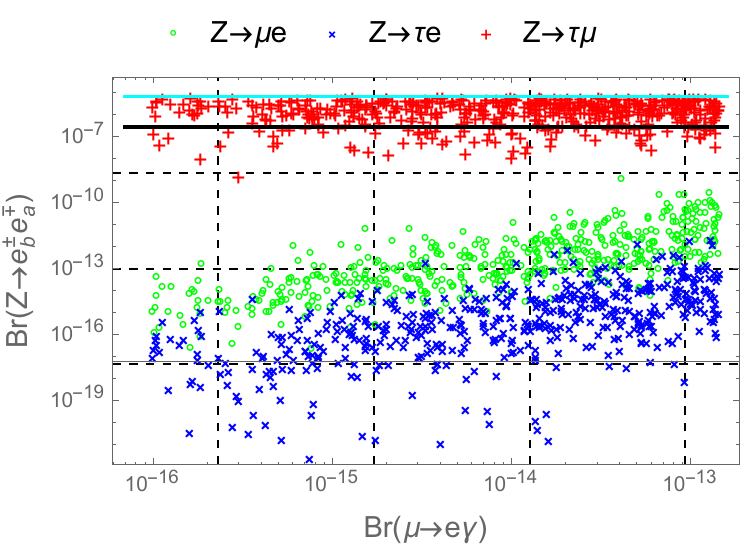} 
\end{tabular}
	\caption{The relationship between decay rates of LFV$Z$ with respect  to $\Delta {a_\mu}$ deviation of muon (left-panel), and decay rates of $\mu\to e\gamma$ channel (right-panel). In each panel, the two cyan and black horizontal lines denote the current experimental upper bounds in Eq.~\eqref{LFV_exp}. The corresponding values are $6.5\times 10^{-6}$ for $\mathrm{Br}(Z\to \tau^\pm\mu^\mp)$ and $2.62\times 10^{-7}$ for $\text{Br}(Z\to \mu^\pm e^\mp)$, respectively.}\label{fig_LFVZeba}
\end{figure}

Next, Fig.~\ref{fig_LFVZeba} shows the correlations between the Br of LFV$Z$ decays, the muon deviation $\Delta a_\mu$ (left panel), and the branching ratio of the radiative decay $\mu\to e\gamma$ (right panel).
where the correlations of Br$(\mu \to e\gamma)$ vs. Br$(Z\to e_b^{\pm}e_a^{\mp})$, in general, have the same properties discussed for the LFV$h$ decay channels, but seems weaker because various LFV$Z$ decay amplitudes consist of complicated parts apart from $f_{ba}$.  This implies that $\mathrm{Br}(Z\to\tau^\pm e^\mp)$ and $\mathrm{Br}(Z\to\mu^\pm e^\mp)$ cannot reach the current experimental upper bounds within the framework of the 331LQ model. Besides, the right panel further shows that $\mathrm{Br}(Z\to\mu^\pm e^\mp)$ is strongly correlated with $\mathrm{Br}(\mu\to e\gamma)$, indicating that future improvements in the experimental sensitivity to the radiative decay $\mu\to e\gamma$ will further constrain the corresponding to the LFV$Z$ decay. 
 As shown in the left panel, although $\mathrm{Br}(Z\to\tau^\pm\mu^\mp)$ exhibits only a weak dependence on $\Delta a_\mu$, it can still reach the current experimental upper bound of $6.5\times10^{-6}$~\cite{ATLAS:2021bdj,ParticleDataGroup:2024cfk}. In contrast, the other two LFV $Z$ decay channels show a slight decreasing trend as $\Delta a_\mu$ approaches the order of $10^{-9}$, which is compatible with the latest discrepancy between the  experimental measurement and the SM prediction based on lattice-QCD calculations~\cite{Aliberti:2025beg}.
 
\begin{figure}[ht]
 	\centering
  	\begin{tabular}{cc}
  			\includegraphics[width=8cm]{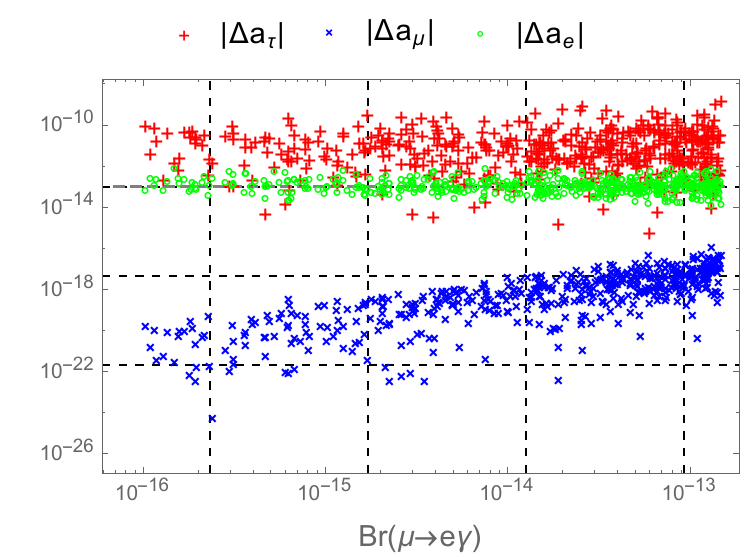} 		&\includegraphics[width=8cm]{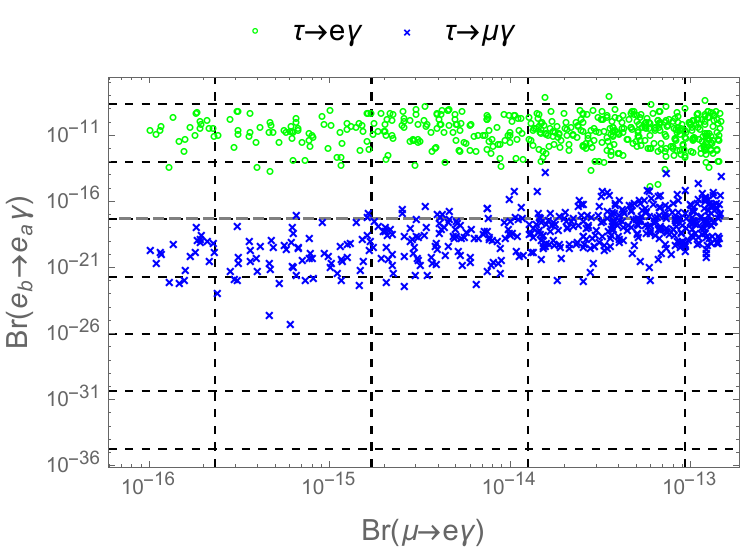}\\
 		\includegraphics[width=8cm]{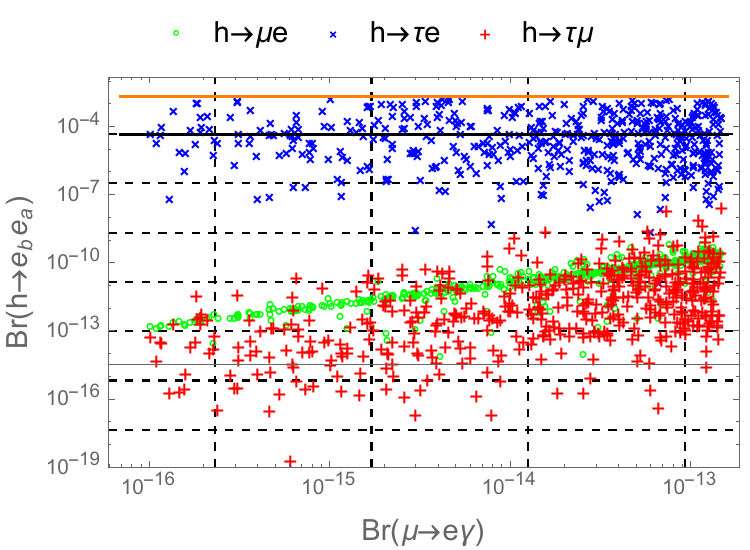} &
 		\includegraphics[width=8cm]{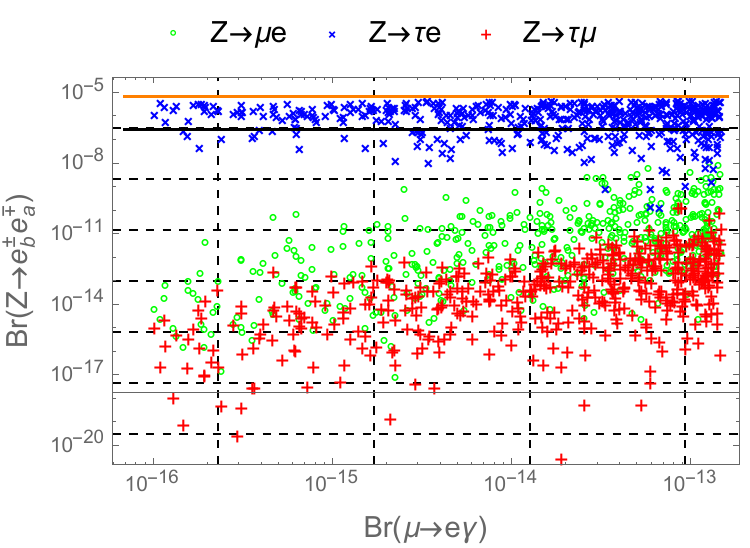} 
 	\end{tabular}
 	\caption{The relationship between $|\Delta a_{e_a}|$ and LFV decay rates  with respect to Br$(\mu \to e\gamma)$. In two bottom panels, the two orange and black horizontal lines denote the current experimental upper bounds in Eq.~\eqref{LFV_exp}. In the left-panel, the corresponding values are $2.0\times 10^{-3}$ for $\text{Br}(h\to\tau e)$ and $4.4\times 10^{-5}$ for $\text{Br}(h\to\mu e)$, respectively. In the right-panel, the corresponding values are $6.5\times 10^{-6}$ for $\text{Br}(Z\to \tau^\pm\mu^\mp)$ and $2.62\times 10^{-7}$ for $\text{Br}(Z\to \mu^\pm e^\mp)$, respectively. }\label{fig_geLFV}
 \end{figure}

For completeness, we consider the regions of parameter space allowing sizable $10^{-14}\leq |\Delta a_{e}| \leq 8\times 10^{-13}$, which still satisfy the current experimental constraints, see illustrations in Fig. \ref{fig_geLFV}.
In addition, $\Delta a_{\mu}$ is now free from below, and all LFV decay rates still satisfy the current experimental sensitivities. We confirm that the properties given in Eq. \eqref{eq:LFVreduced} are still unchanged. On the other hand, $\Delta a_{\mu} \leq \mathcal{O}(10^{-16})$  to keep Br$(\mu\to e\gamma) \leq 1.5\times 10^{-13}$. In addition, only three decay rates of $\mu \to e\gamma$, $h\to\tau e$, and $Z\to \tau^\pm e^{\mp}$ can reach the current and incoming experimental sensitivities. While Br$(Z\to \mu^\pm e^{\mp})\leq 10^{-8}$, Br$(Z\to \tau^\pm \mu^{\mp})\leq 1.5\times 10^{-10}$ are close to the sensitivity of FCC-ee experiment \cite{ATLAS:2021bdj, Dam:2018rfz, FCC:2018byv}, the remaining LFV decay rates are suppressed with the following upper bounds: Br$(\tau \to \mu \gamma)\leq \mathcal{O}(10^{-14})$,  Br$(h \to \mu e)\leq \mathcal{O}(10^{-10})$, Br$(h \to \tau \mu)\leq \mathcal{O}(10^{-8})$.

\section{\label{conclusion} Conclusion}
In this work, we have investigated the phenomenology of the 3-3-1 model extended by a scalar singlet leptoquark. Additionally, we have presented the complete {one-loop contributions from this leptoquark} to LFV decay rates via the PV functions. This representation facilitates numerical calculations and provides a transparent framework for verifying the cancellation of divergences, and we obtained several interesting results. Unlike Ref.~\cite{Doff:2024cap}, which mainly focused on the contributions of the singlet leptoquark to the muon AMM and the radiative decay $\mu\to e\gamma$, {the present work provides numerical investigations of various LFV decays, namely, $\tau\to \mu \gamma, e\gamma$, LFV$h$ and $Z$. We also discuss in detail the correlations between them as well as $\Delta_{e,\mu}$.} After performing an extensive scan over the allowed parameter space, we identified regions compatible with current experimental constraints.  {We found that the singlet leptoquark cannot simultaneously generate sizeable one-loop contributions to both $\Delta a_{e,\mu}$ close to the experimental constraints.}  In addition, several LFV observables, including $\mathrm{Br}(\tau\to e\gamma)$, $\mathrm{Br}(\mu\to e\gamma)$, $\mathrm{Br}(h\to \tau e, \tau\mu)$ and $\mathrm{Br}(Z\to \tau^\pm e^\mp, \tau^\pm\mu^\mp)$, can approach their current experimental upper limits, whereas the remaining decay channels are predicted by 331LQ model to remain several orders of magnitude below the existing bounds.

A remarkable feature of our analysis is the strong correlations among the cLFV and LFV$h$, LFV $Z$ decay channels. In particular, the branching ratios $\mathrm{Br}(h\to\tau\mu)$, $\mathrm{Br}(h\to\tau e)$ and $\mathrm{Br}h\to\mu e)$ exhibit pronounced correlations with their corresponding radiative decays, namely $\tau\to\mu\gamma$, $\tau\to e\gamma$, and $\mu\to e\gamma$, respectively. A similar behavior is observed for the LFV$Z$ decays. These correlations imply that future improvements in the experimental sensitivities to cLFV decays will directly impose more stringent constraints on the corresponding LFV$h$ and LFV$Z$ boson decays, thereby significantly enhancing the predictive power of the model.

Overall, our results demonstrate that the scalar singlet leptoquark provides a viable framework for accommodating the current $(g-2)_\mu$ discrepancy while simultaneously yielding rich and testable phenomenology in cLFV. Future precision measurements of {$(g-2)_{e_a}$} of charged leptons, together with forthcoming searches for cLFV, LFV$h$ and LFV$Z$ decays, will provide powerful probes of the parameter space explored in this work.

\section*{Acknowledgments}
This research is funded by Vietnam National Foundation for Science and Technology Development (NAFOSTED) under the grant number 103.01-2025.04.

\appendix

\section{\label{app:higgs} Higgs bosons}
In this work, the Higgs potential is 
\begin{align}
\label{eq_Vh}
V_{\mathrm{higgs}} =& V^{331}_\Phi + V^{LQ}_S,\crn
V^{331}_\Phi =& \sum_{\Phi} \left[ \mu_\Phi^2 \Phi^{\dagger}\Phi +\lambda_\Phi \left(\Phi^{\dagger}\Phi\right)^2 \right]  + \lambda_{\eta\rho}(\eta^{\dagger}\eta)(\rho^{\dagger}\rho)
+\lambda_{\eta\chi}(\eta^{\dagger}\eta)(\chi^{\dagger}\chi)
+\lambda_{\rho\chi}(\rho^{\dagger}\rho)(\chi^{\dagger}\chi)  \crn
& +\tilde{\lambda}_{\eta\rho} (\eta^{\dagger}\rho)(\rho^{\dagger}\eta) 
+\tilde{\lambda}_{\eta\chi} (\eta^{\dagger}\chi)(\chi^{\dagger}\eta)
+\tilde{\lambda}_{\rho\chi} (\rho^{\dagger}\chi)(\chi^{\dagger}\rho) +\sqrt{2}  f \left(\epsilon_{ijk}\eta^i\rho^j\chi^k +\mathrm{h.c.} \right), \crn
V^{LQ}_S =& \mu^2_S S^*S +\lambda_S\left(S^* S\right)^2+ \left(S^* S\right) \left[\lambda_{S \eta} \eta^{\dagger} \eta+\lambda_{S \rho} \rho^{\dagger} \rho +\lambda_{S \chi} \chi^{\dagger} \chi\right], 
\end{align}
where $\Phi=\eta,\rho,\chi$, and $f$ has a dimension of mass.  In addition,  $	V^{331}_\Phi $ were introduced previously \cite{Diaz:2003dk,Chang:2006aa}, which respects the new general lepton number. The detailed calculations for physical Higgs spectrum in the 331RHN were presented in  previously \cite{Hong:2024swk, Hong:2024yhk,Hong:2022xjg}, in which the leptoquark $S$ plays a role as a charged scalar which does not mix with singly {charged Higgs bosons.} The neutral Higgs states are consistent with Refs.  \cite{Long:1997vbr, Diaz:2003dk,  Ninh:2005su,Hue:2015fbb,Pinheiro:2022bcs}, confirming the existence of a SM-like Higgs boson.  We summary here the main result, using notations given in Ref. \cite{Hong:2024yhk}, combining with the leptoquark part. First, we pay attention to the contributions to the SM-like Higgs boson of two neutral components  $\eta^0_1=(v_{\eta} +R_{\eta} + i  I_{\eta})/\sqrt{2}$, and $ \rho^0=(v_{\rho} +R_{\rho} + i  I_{\rho})/\sqrt{2}$: 
\begin{equation}\label{eq:SMh}
R_{\eta}=s_{\beta} \times  h -c_{\beta} \times h^0_2,\; R_{\rho}= c_{\beta} \times h +s_{\beta} \times h^0_2. 
\end{equation}
The leptoquark $S$ does not mix with all other Higgs boson because of the charged conversation. The respective mass derived from the Higgs potential \eqref{eq_Vh} as follows 
\begin{equation}\label{eq:mS2}
m_S^2=\frac{1}{2} \left(\lambda _{S\eta} v_{\eta}^2+\lambda _{S\rho} v_{\rho}^2+\lambda _{S\chi} v_\chi^2+2 \mu _S^2\right).
\end{equation}
The Lagrangian  parts for the triple couping of the SM-like Higgs boson with leptoquark is: $\mathcal{L}^{hSS} =-V_h= -\lambda_{hSS}hSS^*$ with  $\lambda_{hSS}= { v \left(s_{\beta }^2\lambda _{S\eta}+c_{\beta }^2\lambda _{S\rho }\right)}$. The results show that the leptoquark mass and its couplings with the SM-like Higgs boson is completely independent to the 3-3-1 part of the Higgs potential.

The quartic couplings $\lambda_{S\eta}$ and $\lambda_{S\rho}$ are constrained by vacuum stability conditions and perturbativity requirements. Since these couplings contribute directly to the trilinear vertex $hSS$, they can affect the Higgs phenomenology at the one-loop level. Following the copositivity criteria derived in Ref.~\cite{Kannike:2012pe}, the mixed quartic couplings satisfy
\begin{equation}
\lambda_{S\eta} >-2\sqrt{\lambda_S\lambda_\eta},\, \lambda_{S\rho} >-2\sqrt{\lambda_S\lambda_\rho},
\end{equation}
with $\lambda_\eta,\,\lambda_\rho,\,\lambda_S >0$ these are necessary and sufficient stability conditions \cite{DEramo:2020sqv}. In framework, we chose $-5.0 \leq \lambda_{S\eta}, \lambda_{S\rho} \leq 5.0$ for our numerical investigation, which still matches the perturbation theory.


\end{document}